\documentclass[10pt]{article}

\usepackage{amsmath}
\usepackage{amssymb}
\usepackage{graphicx}
\usepackage{natbib}
\usepackage{color}
\usepackage{url}
\usepackage{setspace}
\usepackage{authblk}
\usepackage{lineno}

\usepackage{setspace}

\newcommand{\sed}[1]{{\bf \textcolor{blue}{#1}}}
\renewcommand{\sed}[1]{#1}

\renewcommand{\vec}[1]{\boldsymbol{#1}}

\newcommand{\del}{\partial}

\title{The Influence of Galactic Cosmic Rays on Ion-Neutral Hydrocarbon Chemistry in the Upper 
Atmospheres of Free-Floating Exoplanets}
\author{P. B. Rimmer\thanks{email:~pr33@st-andrews.ac.uk}~}
\author{Ch. Helling}
\affil{SUPA, School of Physics and Astronomy, University of St Andrews,
St Andrews, KY16 9SS, UK}
\author{\sed{C. Bilger}}
\affil{\sed{Department of Engineering, University of Cambridge, Cambridge, CB2 1PZ, UK}}
\date{Keywords: planets and satellites: atmospheres,
cosmic rays, 
atmospheric effects, 
astrochemistry}

\begin{document}

\maketitle

\begin{abstract}
 Cosmic rays may be linked to the formation of volatiles necessary for prebiotic chemistry. \sed{We} explore the 
effect of cosmic rays in a hydrogen-dominated atmosphere, \sed{as a proof-of-concept that ion-neutral chemistry may
be important for modelling hydrogen-dominated atmospheres}. \sed{In order to accomplish this, we} utilize Monte Carlo cosmic ray 
transport models with particle energies of $10^6$ eV $< E < 10^{12}$ eV in order to investigate the cosmic ray 
enhancement of free electrons in substellar atmospheres. \sed{Ion-neutral chemistry is then} applied to 
a \textsc{Drift-Phoenix} model of a free-floating giant gas planet. Our results suggest that the activation of ion-neutral 
chemistry in the upper atmosphere significantly enhances formation rates for various species, and we find that C$_2$H$_2$, 
C$_2$H$_4$, NH$_3$, C$_6$H$_6$ and \sed{possibly} C$_{10}$H are enhanced in the upper atmospheres because of cosmic rays. 
\sed{Our results suggest a potential connection between cosmic ray chemistry and the hazes observed} in the upper atmospheres 
of \sed{various} extrasolar planets. Chemi-ionization reactions are briefly discussed, \sed{as they} may enhance 
the degree of ionization in the cloud layer.
\end{abstract}


\section{Introduction}
\label{sec:intro}

Life requires a considerable variety of chemical ingredients, and these ingredients are composed
mostly of four elements: hydrogen, carbon, nitrogen and oxygen. The backbone species for prebiotic
chemistry include the reactive species HCHO, HCN, ethylene, cyanoacetylene, and acetylene 
\citep{Miller2006}, and it is an open question how or even where these species were first
formed, whether in the interstellar medium \citep{Hoyle2000}, on the backs of meteorites, 
in the atmosphere, or deep within the oceans of the archaic earth \citep{Orgel1998}. Energetic 
processes, such as photolysis and cosmic ray ionization affect atmospheric and geological chemistry
on earth, and may have an important role to play in the formation of prebiotically relevant radical 
species. The atmosphere of Jupiter and Jupiter-like planets provide a natural laboratory in which to 
study the effects energetic processes have on hydrocarbon and organic chemistry. The atmospheres 
of giant gas planets are hydrogen dominated and have significant elemental concentrations of carbon 
and oxygen \citep{Lodders2004}, and many of the same energetic processes take place in their atmospheres 
as on earth, such as photodissociation \citep{Moses2005}, cosmic ray ionization \citep{Whitten2008} and 
lightning \citep{Gurnett1979}.

There has been significant work on cosmic ray transport and ionization in the atmosphere of Titan. 
The effect cosmic rays have on ionizing and dissociating carbon and nitrogen species 
\citep{Capone1980,Capone1983,Molina1999}, enhancing the formation of aerosols \citep{Sittler2010}, 
and the role of cosmic rays in instigating lightning \citep{Borucki1987} have been explored. 
Cosmic ray transport in Titan has also been investigated \citep[for example,][]{Molina1999b}. 
In terms of electron production, the transport model for Titan bears qualitatively similar results 
to the cosmic ray transport models on Earth \citep[e.g.,][]{Velinov2009}. If results are cast in 
terms of column density, both seem to be well-fit by a Poisson distribution and peak at roughly the 
same atmospheric depth ($\sim 100$ g cm$^{-2}$). Initial studies on cosmic ray ionization in 
Jupiter \citep{Whitten2008} and exoplanets \citep{Rimmer2013} support this result.

The effects of cosmic ray ionization on the chemistry \sed{of} extrasolar planet atmospheres has not yet 
been investigated, although \cite{Moses2011} speculate that cosmic ray ionization may be connected 
to the hazes in the upper atmosphere of HD 189733b observed by \citet{Pont2008} and \citet{Sing2011}.
An investigation into hydrocarbon chemistry in the atmospheres of extrasolar giant gas planets is
interesting from an astrobiological perspective because such a study will provide insight into 
the significance of different initial conditions and physical parameters on this prebiotic chemistry.
\sed{We present a proof-of-concept on the potential effect of cosmic rays in the atmosphere between 
10 $\mu$bar and 1 nbar. In order to accomplish this} we take the results of cosmic ray ionization on a 
\textsc{Drift-Phoenix} model atmosphere of a Jupiter-like planet from \citet{Rimmer2013}, and apply them 
to an astrochemical network adapted to the atmospheric environment of a Jupiter-like planet. A brief 
description of \textsc{Drift-Phoenix} and the cosmic ray transport is provided in Section \ref{sec:crt}. 
We then discuss the chemical network and its results in \ref{sec:model}. Section \ref{sec:conclusion} 
contains our conclusions.

\section{The Model Atmosphere and Cosmic Ray Transport}
\label{sec:crt}

\begin{figure}
\centering
\includegraphics[width=0.8\columnwidth]{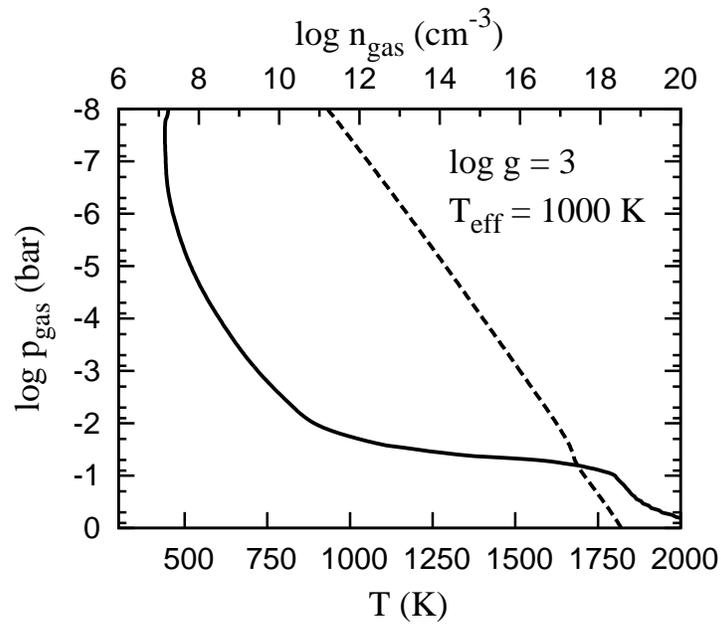}
\caption{\sed{The temperature, $T$ [K] (solid,bottom) and density, $\log n_{\rm gas}$ [cm$^{-3}$] (dashed,top) plotted as a 
function of pressure, $\log p_{\rm gas}$ [bar], for our model} \textsc{Drift-Phoenix} \sed{atmosphere,
when $\log g = 3$, $T_{\rm eff} = 1000$ K, and at solar metallicity.}}
\label{fig:pT}
\end{figure}

The chemical calculations in Section \ref{sec:model} are built upon a model atmosphere of 
\textsc{Drift-Phoenix} \citep[from][]{Dehn2007,Helling2008,Witte2009}, and cosmic ray transport calculations 
from \citet{Rimmer2013}. \textsc{Drift-Phoenix} is a model that self-consistently combines the 
calculation of hydrostatic equilibrium and comprehensive radiative transfer for a substellar object or 
warm exoplanet with a dust formation model that incorporates seed nucleation, growth, gravitational 
settling, and evaporation of dust grains. This self-consistent approach has the advantage of capturing 
the effect of the dust on the gas-phase, e.g. the elemental \sed{depletion} and the backwarming effect of 
the dust on the gas-phase temperature. \textsc{Drift-Phoenix} has been utilized in explorations of 
non-thermodynamic charging of the atmosphere by, e.g. dust-dust collisions and Alfv\'{e}n ionization 
\citep{Helling2011b,Stark2013}. This model takes as input elemental abundances, effective temperature 
($T_{\rm eff}$, K) and surface gravity ($\log g$, with $g$ in units of cm s$^{-2}$). The \sed{example} atmosphere 
we consider \sed{here} is that of a model free-floating exoplanet, with $\log g = 3$, $T_{\rm eff} = 1000$ K
and solar metallicity. \sed{The pressure-temperature structure of the model is plotted in Figure \ref{fig:pT}.}

We apply a Monte Carlo model to determine cosmic ray transport for cosmic rays of energy $E < 10^9$ eV.
In this model, detailed in \citet{Rimmer2012}, we take 10000 cosmic rays and propagate them through a 
model exosphere \citep[][their Sec. 2]{Rimmer2013} and \textsc{Drift-Phoenix} atmosphere. The
cosmic rays travel a distance such that $\sigma \Delta N \ll 1$ ($\sigma$ [cm$^2$] denotes the 
collisional cross-section between the cosmic ray and a gas-phase atmospheric species, 
and $\Delta N$ [cm$^{-2}$] is the column density of the distance traveled). A fraction of the cosmic
rays will have experienced a collision and will lose some of their energy, $W(E)$ [eV]. Alfv\'{e}n
wave generation by cosmic rays \citep{Skilling1976} is also accounted for.

\sed{An analytical form for the column-dependent ionization rate, $Q$, as a function of column 
density, $N_{col}$ [cm$^{-2}$] from \citet{Rimmer2013} is:
\begin{align}
 Q & (N_{col}) =  Q_{\rm HECR}(N_{col}) \notag\\ 
& +\zeta_0 n_{\rm gas} \times \begin{cases}480 & \!\!\!\!\! \text{if } N_{col} < N_1 \\
   1 + (N_0/N_{col})^{1.2} & \!\!\!\!\!  \text{if } N_1 < N_{col} < N_2\\
  e^{-kN_{col}} & \!\!\!\!\!  \text{if } N_{col} < N_2\\
 \end{cases}
 \label{eqn:e-production}
\end{align}
where $\zeta_0 = 10^{-17}$ s$^{-1}$ is the standard ionization rate 
in the dense interstellar medium, and the column densities $N_0 = 7.85 \times 10^{21}$ 
cm$^{-2}$, $N_1 = 4.6 \times 10^{19}$ cm$^{-2}$, 
$N_2 = 5.0 \times 10^{23}$ cm$^{-2}$, and $k = 1.7 \times 10^{-26}$ cm$^2$ 
are fitting parameters. The value for the high-energy cosmic ray ionization rate, $
Q_{\rm HECR}(N_{col})$ is given in \citet[][their Fig. 5]{Rimmer2013}.}

 For cosmic rays with $E > 10^9$ eV, the cosmic ray transport is treated by 
the analytical form of \citet{Velinov2008,Velinov2009}, appropriately modified for our model atmosphere.
\citet{Rimmer2013} provide a detailed descriptions of the modifications to the calculations of 
\citet{Velinov2008} and \citet{Rimmer2012}.

\section{Gas-Phase Chemical Kinetics Model}
\label{sec:model}

Now that the rate of electron production by galactic cosmic rays is determined, we study the effect 
that this rate will have on the atmospheric gas by applying a gas-phase network. Our chemical model 
allows us to explore first findings with respect to the local hydro-carbon chemistry. The electrons
in the atmosphere will be freed by cosmic rays and by the secondary photons produced from their 
interaction with the gas. Secondary photons are generated via the Prasad-Tarafdar mechanism 
\citep{Prasad1983}. We utilize the detailed emission spectrum from \citet{Gredel1989}. Photoionization 
and photodissociation rates for secondary photons are included in the model. The rate of collisional 
de-excitation varies throughout the atmosphere. At the upper limit for our model, $T_{\rm gas} = 800$ K,
the $\rm{H_2}$ number density is $\sim 10^{14}$ cm$^{-3}$ for our model giant gas planet atmosphere. 
The cross-section for collisional de-excitation is on the order of $\sigma_{\rm coll} \sim 10^{-23} 
- 10^{-22}$ cm$^{2}$ \citep{Shull1978}, corresponding to a time-scale for collisional de-excitation of 
$t_{\rm coll} \approx \big( n_{\rm gas} \sigma_{\rm coll} v_{\rm coll} \big)^{-1} \sim 100$ s, where 
$v_{\rm coll}$ is the average collisional velocity of the gas at 800 K. This is much longer than
the time-scale for spontaneous emission \citep{Prasad1983}, and so we ignore the effect of collisional 
de-excitation on our emission spectrum. \citet{Millar1997,McElroy2013} claim that the photoionization 
and photodissociation rates for secondary photons are valid over a temperature range of 
$10$ K $ < T_{\rm eff} < 41000$ K.

Free electrons will also be destroyed by recombination processes, e.g. with ions. We compare these two 
competing processes in detail: If the cosmic ray ionization rate is much greater than the recombination 
rate, then we can expect a significant electron enhancement from cosmic rays. If, however, the 
recombination is much more rapid than the ionization, then the number of free electrons will remain 
low. We examine these two rates by concentrating on the number of free electrons produced by cosmic 
rays and cosmic-ray photons. We utilize a self-consistent gas-phase time-dependent chemical model that 
includes various cosmic ray, secondary electron and secondary photo-ionization chemical processes,
and recombination reactions (described below). This model is calculated iteratively at different 
depths with the value of $Q$ [cm$^{-3}$ s$^{-1}$] calculated using \citet[][their Eq. 23]{Rimmer2013}. 
As result of this procedure, we determine the number of free electrons and chemical abundances as a 
function of the local atmospheric pressure for our model giant gas planet atmosphere.

\subsection{The kinetic gas-phase chemistry model}
Various chemical kinetics networks have been developed to model the non-equilibrium atmospheric chemistry 
of giant gas planets \citep{Moses2000,Zahnle2009b,Venot2012}. \citet{Bilger2013} introduce a 
new approach to non-equilibrium chemistry that explores individual and immediate destruction reactions
for various species absent from the gas-phase in the upper atmosphere. These models involve robust 
chemical networks and calculate radiative transfer and vertical mixing, but they do not account for 
ion-neutral chemistry. We thus have adapted an interstellar chemical model with a detailed cosmic ray 
chemistry in order to calculate the impact of cosmic rays on the degree of ionization by chemical 
kinetic processes. Molecular hydrogen dominates throughout our model atmospheres, and interstellar 
kinetics models contain a detailed treatment of the cosmic ray ionization of H and H$_2$, among other 
species, as well as recombination rates for all ions included in the networks.

The density profile and equilibrium chemical abundances for the atmosphere, from the exobase and through 
the cloud layer, are taken from the \textsc{Drift-Phoenix} model atmosphere results. The chemical 
kinetics are modeled using the \textsc{Nahoon} gas-phase time-dependent astrochemical model 
\citep{Wakelam2005,Wakelam2012}. The newest version of this code can model objects with temperatures of 
$T \lesssim 800$ K \citep{Harada2010}. We use the high-temperature OSU 2010 Network from 
\url{http://physics.ohio-state.edu/~eric/research_files/osu_09_2010_ht}.\\ This network comprises 461 
species and well over 4000 reactions, including cosmic ray ionization, photo-ionization, 
photo-dissociation, ion-neutral and neutral-neutral reactions. The additional reactions from the 
high-temperature network of \citet{Harada2010} include processes with activation energies of 
magnitude $\sim 2000$ K, reverse reactions and collisional dissociation.

We effectively apply a model at our model atmosphere for the chemical composition at different 
atmospheric heights. For irradiated atmospheres and atmospheres of rotating bodies, vertical mixing can 
be significant, especially in the upper atmosphere. For non-rotating planets far from their host star, 
however, vertical mixing timescales are on the order of $10^3 - 10^7$ years 
\citep[][esp. their Fig. 2]{Woitke2004}, and are orders of magnitude longer than the chemical 
time-scales over the entire range of atmospheric pressures. We therefore expect that the atmospheric 
chemistry can effectively be treated as a series of zero-dimensional chemical networks. For irradiated
atmospheres, vertical mixing time scales can be much shorter \citep{Showman2009,Moses2011}. \sed{For
initial abundances at each height, we use the thermodynamic equilibrium abundances calculated in the manner
described by \citet{Bilger2013}, where available. If the thermochemistry is not
calculated for a given species in the OSU high Temperature network, we set the initial abundances equal to
zero. The model is calculated until it reaches steady-state. We have not yet explored the effect of varying 
the initial abundances.}

\sed{Because our model is applied to a hypothetical free-floating giant gas planet, vertical mixing 
and UV photoionization need not be accounted to approximate the non-equilibrium chemistry, because these processes are
expected to be negligible in such an environment. This is appropriate for a first proof-of-concept, but
it does mean that our methods cannot be directly applied to irradiated planets, where vertical mixing
time-scales are short and UV photoionization and photodissociation has a major effect on chemistry in
the upper atmosphere. We plan to apply an expanded version of our ion-neutral network to a Hot Jupiter 
in a future paper.}

\sed{Our model does not account for termolecular neutral-neutral or ion-neutral reactions. At some
atmospheric density, termolecular reactions will become important. In order to estimate when this is,
we follow the example of \citet{Aikawa1999} and calculate the ``typical'' rate for a bimolecular 
reaction ($k_2 \sim 10^{-11}$ cm$^6$ s$^{-1}$) divided by a ``typical'' rate for a termolecular reaction, 
$k_3 \sim 10^{-30}$ cm$^6$ s$^{-1}$):
\begin{equation}
 \dfrac{k_2}{k_3} \sim 10^{19} \; {\rm cm^{-3}}.
\end{equation}
Our critical density below which three-body neutral-neutral reactions are not significant would be 
$n_c \sim 10^{17}$ cm$^{-3}$. Alternatively, \citet{Woods2007} estimate that $n_c \sim 10^{14}$ cm$^{-3}$.
If we use the lower value, then the critical pressure, above which termolecular reactions dominate is 
$p_c \approx 10^{-5}$ bar. If we use the higher value of $\sim 10^{17}$ cm$^{-3}$, then the cutoff moves
to higher pressure, of about $p_c \approx 0.1$ bar. We use the stronger limit of $n_c \sim 10^{14}$ cm$^{-3}$.}

\textsc{Nahoon} treats the kinetic chemistry as a series of rate equations. A given rate equation for a 
species $A$ formed by $B + C \rightarrow A + X$, and destroyed by $A + D \rightarrow Y + Z$, is of the 
form:
\begin{equation}
 f_1 = \dfrac{dn(A)}{dt} = k_{BC} n(B) n(C) - k_{AD} n(A) n(D) v_{\rm coll}  + \cdots,
\end{equation}
where $k_{BC}$, $k_{AD}$ are the rates of formation and destruction of $A$, respectively. 
We can represent the rate equation for the $i$-th species as $f_i$, $i$ ranging from $1$ to the total 
number of species, $N$. This system of equations has the Jacobian:
\begin{equation}
 J = \begin{pmatrix}
      \del f_1/\del n_1 & \cdots & \del f_1/\del n_N\\
         \vdots & \ddots & \vdots \\
      \del f_N/\del n_1 & \cdots & \del f_N/\del n_N
     \end{pmatrix}
\end{equation}
where $N$ is the total number of gas-phase species.

\begin{figure}
\centering
\includegraphics[width=0.8\columnwidth]{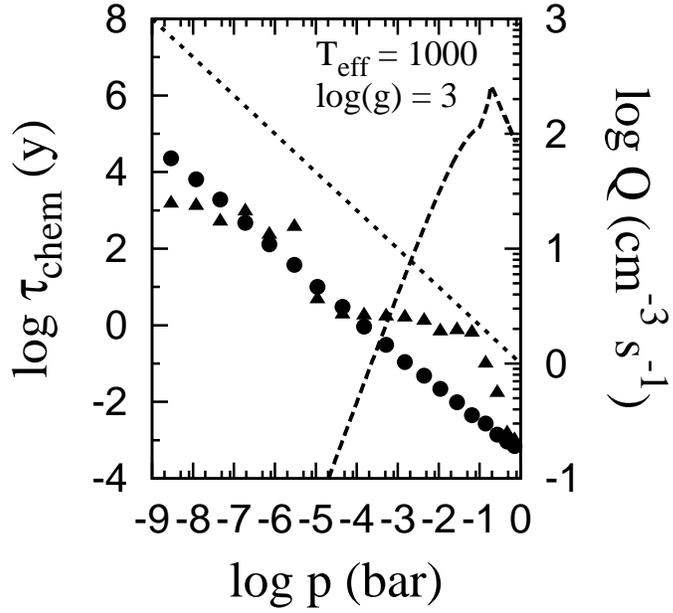}
\caption{The chemical relaxation time-scale toward steady state, $\tau_{\rm chem}$, as a function 
of the local gas pressure, for the giant gas planet model atmosphere 
($T_{\rm eff} = 1500$, $\log g = 3$, solar metallicity).
Each data point represents the time for the \textsc{Nahoon} model to achieve steady state, with cosmic 
rays (triangles, left axis) and without cosmic rays (circles, left axis). The dotted line (left axis)
represents the total runtime for the \textsc{Nahoon} model. The divergence from $1/n_{\rm gas}$ in 
run-times is probably due to the higher number of free electrons produced by cosmic rays. The
cosmic ray electron production rate, $Q$ [cm$^{-3}$ s$^{-1}$], is also shown (dashed line, right axis).}
\label{fig:time}
\end{figure}

 We run this model for the whole temperature-density profile for the model atmosphere considered here.
 We allow the network to evolve to steady state, and the time the system takes to reach steady state 
depends on the atmospheric depth. The time-steps decrease in proportion to the increasing total density. 
The relaxation time-scale for the model can be determined from the inverse of the Jacobian of the 
system taken when all the species in the system have reached a steady state: $n_i = \mathring{n}_i$ for 
a given species, $i$, and $\mathring{f}_i \equiv f(\mathring{n}_i) = 0$. Specifically, one can determine 
the relaxation time-scale, $\tau_{\rm chem}$, to achieve steady state to be equal to the inverse of the 
eigenvalue, $\lambda_i$, with the smallest absolute real component, or 
\citep[][their Eq. 117]{Woitke2009}:
\begin{equation}
 \tau_{\rm chem} = {\rm max}|{\rm Re}\{\lambda_i\}^{-1}|.
\label{eqn:lambda-time}
\end{equation}
For the eigen-vectors of the Jacobian, $\vec{j}$:
\begin{equation}
  J\vec{j} = \lambda_i\vec{j}.
\end{equation}
All $\lambda_i$ must have the same proportionality relationship as the components of $J$, so:
\begin{equation}
 \lambda_i \propto \dfrac{\del f_i}{\del n_i} \propto \dfrac{f_i - \mathring{f}_i}{n_i - \mathring{n}_i}.
\end{equation}
Since the deviation from steady state, $n_i - \mathring{n}_i$, is proportional to the total density of 
the system, $n_{\rm gas} = \Sigma_i n_i$, and the deviation $f_i - \mathring{f}_i = f_i$ is proportional 
to $n_{\rm gas}^2$ for two-body reactions, it follows that:
\begin{equation}
 \tau_{\rm chem} \propto \dfrac{1}{n_{\rm gas}}.
\label{eqn:time-scale}
\end{equation}
We run \textsc{Nahoon} until steady state, $\mathring{t} = \tau_{\rm chem}$, with time-steps 
proportional to $\tau_{\rm chem}$ and therefore to $1/n_{\rm gas}$.  Equation (\ref{eqn:time-scale}) 
does not account for the temperature dependence. Also, first-order reactions, particularly the cosmic 
ray ionization reactions, will have relaxation times that scale differently with $n_{\rm gas}$ \sed{and
this is why the time to equilibrium is different with cosmic ray ionization than without cosmic ray 
ionization. Ionization rate coefficients for certain species are quite small, and this can increase
the time to steady-state.}  
A complete determination of $\tau_{\rm chem}$ therefore would require solving Eq. (\ref{eqn:lambda-time}). 
In practice, however, we find that adjusting the runtime of \textsc{Nahoon} by the proportionality
$1/n_{\rm gas}$ achieves steady state at each point in the atmosphere. Figure \ref{fig:time} contains a 
plot of $\tau_{\rm chem}$ as a function of the atmospheric gas pressure both with and without cosmic ray 
ionization. The figure shows that significantly different amounts of time are required in order to 
reach steady state for different depths. Near the exobase, the time-scale can be on the order of 
$10^3$ years, while in the cloud layer $\tau_{\rm chem}$ is on the order of hours. When cosmic ray 
ionization is included, $\tau_{\rm chem}$ is less than $1/n_{\rm gas}$ by two orders of magnitude in 
the very upper atmosphere, and is greater than $1/n_{\rm gas}$ by more than an order of magnitude within 
the cloud layer. At the cloud base, however, the value of $\tau_{\rm chem}$ with cosmic rays converges 
\sed{to} the value of $\tau_{\rm chem}$ without cosmic rays.

Cosmic ray ionization is included in the OSU network as rates with coefficients, $\alpha_X$, scaling 
the primary ionization rate for the reaction $X$ such that $\zeta_X = \alpha_X \zeta_p$. The 
\textsc{Nahoon} model is iterated toward steady-state at different depths, where the local gas density 
and temperature ($n(z)$, $T(z)$) and the cosmic ray flux density change with depth. External UV fields 
are not considered in this study, in order to clearly identify the cosmic ray contribution to 
ionization. The impact of photons generated by the secondary electrons, so-called cosmic ray photons, 
is included, however.

\sed{The results of our network provide some interesting predictions for free-floating giant gas planets.
Some of these results may be applicable also to irradiated exoplanets, although the enhanced mixing and
photochemistry may drastically change some of these predictions. First, our results predict where the free
charges, both positive and negative, are likely to reside, chemically. Our model predictions for the most 
abundant cations and anions are presented in Section \ref{sec:cations}. Because the focus of this paper is on
prebiotic chemistry, and because of the strong effect cosmic rays have upon them, complex hydrocarbons are the 
neutral species of most interest. Our results for complex hydrocarbons are presented in 
Section \ref{sec:carbon-chem}. In this section, we also discuss a neutral-neutral reaction that produces 
free electrons, and will speculate on its effect on the degree of ionization within a brown dwarf atmosphere.
Section \ref{sec:ammonia} concludes our discussion on chemistry with ammonia.}

\subsection{\sed{Atmospheric Cations and Anions}}
\label{sec:cations}

According to our chemical network, cosmic rays primarily ionize H$_2$, H$_2$O, C, and He, the most 
abundant neutral species in the upper atmosphere of an oxygen-rich, hydrogen-dominated planet. 
The ions H$_2^+$, H$_2$O$^+$, C$^+$, and He$^+$ do not retain the bulk of the excess positive charge 
for long, but react away or exchange their excess charge.

\begin{figure}
\centering
\includegraphics[width=0.8\columnwidth]{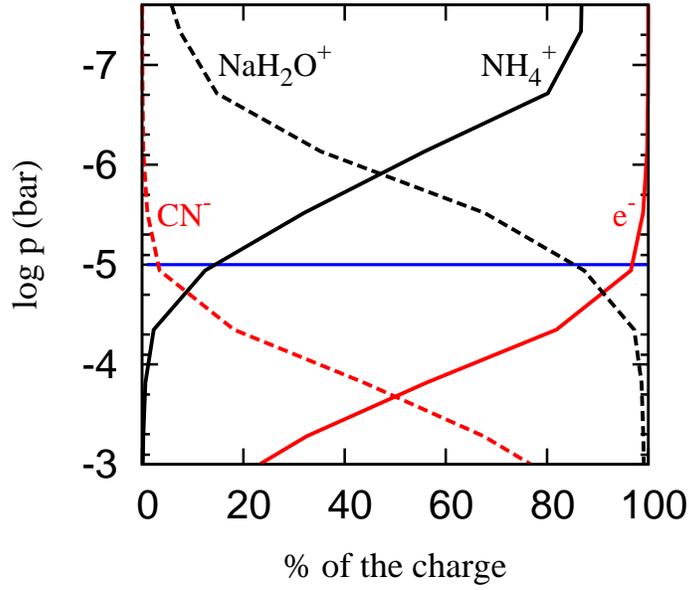}
\caption{ The percentage of positive (black) and negative (red) charge contained by a given species 
as a function of pressure, $p$ [bar] for our model giant gas planet atmosphere 
($T_{\rm eff} = 1000$ K, $\log g = 3$). The negative charge is mostly in the form of electrons and 
CN$^-$. The positive charge is carried primarily by NH$_4^+$ in the upper atmosphere and NaH$_2$O$^+$ 
in the cloud layer. The rest of the positive charge is mostly in the form of large carbon-containing 
ions, e.g. ${\rm C_7H_2^+}$, ${\rm C_7H_3^+}$ and ${\rm C_8H_4^+}$. \sed{A blue
horizontal line indicates the pressure above which termolecular reactions may dominate.}}
\label{fig:e-chem}
\end{figure}

We find that for our model exoplanet atmosphere, when $p_{\rm gas} < 10^{-6}$ bar, up to 90\% of 
positive charge is carried by NH$_4^+$, and the rest of the positive charge is carried by various
large carbon-bearing cations. When $p_{\rm gas} > 10^{-6}$ bar, the most abundant cation becomes 
NaH$_2$O$^+$, carrying more than 95\% of the charge for $p_{\rm gas} > 5 \times 10^{-4}$ bar. There are 
two reasons for the transition between these two species: cosmic ray ionization and enhanced charge 
exchange at high densities. Figure \ref{fig:e-chem} shows which species carry most of the positive 
charge.

In the upper atmosphere, NH$_4^+$ is formed primarily by the reactions:
\begin{align}
 {\rm C_3H_2N^+ + NH_3} &\rightarrow {\rm NH_4^+ + HC_3N}, \label{eqn:poly1}\\ 
 {\rm C_5H_2N^+ + NH_3} &\rightarrow {\rm NH_4^+ + HC_5N}, \label{eqn:poly2}\\ 
 {\rm C_7H_2N^+ + NH_3} &\rightarrow {\rm NH_4^+ + HC_7N}, \label{eqn:poly3}\; {\rm and} \\ 
 {\rm H_3O^+ + NH_3} &\rightarrow {\rm NH_4^+ + H_2O}.
\end{align}
Reactions (\ref{eqn:poly1})-(\ref{eqn:poly3}) are also the dominant formation pathways for the 
cyanopolyynes in the upper atmosphere. The large cations C$_{2n+1}$H$_2$N$^+$ as well as H$_3$O$^+$ 
are connected, via H$_3^+$, to cosmic ray driven chemistry. Since the cosmic rays do not penetrate 
through the cloud layer, the C$_{2n+1}$H$_2$N$^+$ ions become depleted, and NH$_4^+$ is no longer 
efficiently formed.

At $p_{\rm gas} > 10^{-6}$ bar, as the gas density increases, collisions become more frequent and charge 
exchange reactions dominate. Atomic sodium collects a large amount of the charge through these charge 
exchange reactions. Then:
\begin{align}
 {\rm Na^+ + H_2} &\rightarrow {\rm NaH_2^+} , \\
 {\rm NaH_2^+ + H_2O} &\rightarrow {\rm NaH_2O^+ + H_2} , \\
 {\rm Na^+ + H_2O} &\rightarrow {\rm NaH_2O^+},
 \end{align}
which proceed rapidly because of the high abundances of H$_2$ and water. NH$_4^+$ and NaH$_2$O$^+$ are 
both destroyed mostly by dissociative recombination. 

\subsection{Cosmic Rays and Carbon Chemistry in the Upper Atmosphere}
\label{sec:carbon-chem}

It has been suggested that cosmic rays enhance aerosol production in the terrestrial atmosphere 
\citep[e.g.][]{Shumilov1996} and that they potentially initiate chemical reactions that allow the 
formation of mesospheric haze layers such as those observed on the exoplanet HD 189733b by 
\citet{Pont2008} and \citet{Sing2011} and those observed on Titan 
\citep{Rages1983,Porco2005,Liang2007,Lavvas2009}. We are interested in the effects that 
cosmic rays can have on the carbon chemistry in extraterrestrial, oxygen-rich atmospheres, such as those
of giant gas planets. Especially relevant are the chemical abundances of the largest carbon species in 
the gas-phase model, which is likely connected to PAH production in planetary atmospheres 
\citep{Wilson2003}. Although our network does not incorporate a complete PAH chemistry, it does contain 
chemical pathways that reach into long carbon chains, the longest chain being ${\rm C_{10}H}$.

\begin{figure}
\centering
\includegraphics[width=0.8\columnwidth]{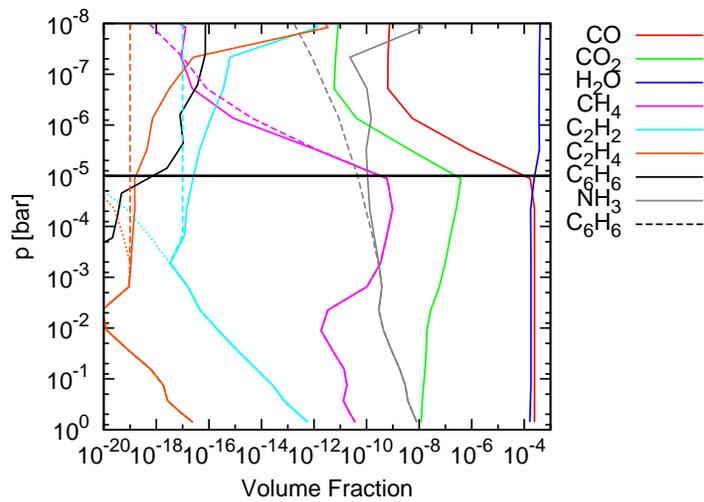}
\caption{\sed{Volume fraction of various species as a function of the gas pressure, $p$ [bar] for the model atmosphere of a 
free-floating giant gas planet ($T_{\rm eff} = 1000$ K, $\log g = 3$), obtained by combining our results to those of
\citet{Bilger2013}. The results of \citet[][dotted]{Bilger2013}, the results assuming chemical quenching of 
C$_2$H$_2$ and C$_2$H$_4$ at height at $\sim 10^{-3}$ (dashed), and the results with cosmic ray ionization (solid) are all 
presented in this plot. A thick black horizontal line indicates the pressure above which termolecular reactions may dominate.} }
\label{fig:chem}
\end{figure}

\sed{The most meaningful information from our chemical model is the enhanced or reduced abundance
of various species due to ion-neutral chemistry, compared to their abundances according to purely neutral-neutral
chemistry. Although this shows us the effect cosmic-ray driven chemistry can have on the abundances of carbon
bearing species, it does not tell us whether these enhancements are observable. An enhancement of 10 orders
of magnitude on a species with a volume fraction, from $10^{-40}$ to $10^{-30}$, is significant, but unobservable.
In order to answer this question, we apply the thermochemical equilibrium results for a $\log g = 3$, $1000$ K 
atmosphere using the calculations from \citet{Bilger2013}, to the degree that the various species have been enhanced or 
reduced. We take the chemical equilibrium number density of a species, $X$, $n_{\rm eq}(X)$ [cm$^{-3}$] and solve for the 
non-equilibrium abundance, $n_{\rm neq}(X)$ [cm$^{-3}$] by:
\begin{equation}
n_{\rm neq}(X) = \xi n_{\rm eq}(X), 
\end{equation}
where $\xi$ is the amount that the abundance is enhanced or reduce by cosmic rays. We plot the predicted volume fractions
of CO, CO$_2$, H$_2$O, CH$_4$, C$_2$H$_2$, C$_2$H$_4$ and NH$_3$, as a function of pressure, in Figure \ref{fig:chem}.
Although C$_2$H and C$_{10}$H are also enhanced by orders of magnitude, the results are not shown in this figure. For
C$_2$H, this is because the resulting volume fraction is lower than $10^{-20}$ throughout the model atmosphere. For
C$_{10}$H, this is because the equilibrium concentration is unknown. The results for cosmic ray enhancement are not reliable
when $p_{\rm gas} > 10^{-5}$ bar because termolecular reactions may begin to take over. When this happens, so long as vertical
mixing timescales are not too small, the added reactions will bring the system to thermochemical equilibrium; equilibrium
abundances are also presented in the figure.}

\subsubsection{\sed{C$_2$H$_2$ and C$_2$H$_4$}}

\sed{For the predicted cosmic ray enhancement of C$_2$H$_2$ and C$_2$H$_4$, we assume chemical quenching at $10^{-3}$ bar,
although it may well be quenched at much higher pressures of $\sim 0.01$ bar \citep{Moses2011}. 
With this assumption, we find that both C$_2$H$_2$ and C$_2$H$_4$ are brought to volume fractions of $\sim 10^{-12}$
when $p_{\rm gas} \approx 10^{-8}$ bar. This is compared to the quenched abundance in the absence of cosmic rays, of 
$\sim 10^{-17}$ for C$_2$H$_2$ and $\sim 10^{-19}$ for C$_2$H$_4$.}

\sed{Carbon monoxide and methane are largely unaffected by cosmic ray ionization,
although methane is depleted by about one order of magnitude at $10^{-8}$ bar. Both methane and carbon monoxide decrease 
rapidly with decreasing pressure when $p_{\rm gas} \lesssim 10^{-6}$ bar and, according to \citet{Bilger2013}, at these 
pressures and temperatures a significant amount of the carbon is atomic. The cosmic ray chemistry does impact methane,
but only in the very upper atmosphere, where it is not very abundant. Cosmic rays deplete the methane by approximately
one order of magnitude, from a volume fraction of $10^{-16}$ to $10^{-17}$.}

\sed{Acetylene and ethylene, as well as C$_2$H, are enhanced by various complex reaction pathways. 
One such pathway to acetylene, dominant at $10^{-8}$ bar, when there is a high fraction neutral carbon, is:
\begin{align}
 {\rm H_2 + CR} &\rightarrow {\rm H_2^+ + CR} + e^-, \\
 {\rm H_2^+ + H_2} &\rightarrow {\rm H_3^+ + H}, \\
 {\rm H_3^+ + C} &\rightarrow {\rm CH^+ + H_2}, \\
 {\rm CH^+ + 3H_2} &\rightarrow {\rm CH_5^+ + 2H}, \\
 {\rm CH_5^+ + C} &\rightarrow {\rm C_2H_3^+ + H_2}, \\
 {\rm C_2H_3^+} + e^- &\rightarrow {\rm C_2H_2 + H}.
\end{align}
There are dozens of pathways from H$_3^+$ to the complex hydrocarbons listed above, and it would be beyond the scope
of this paper to list which multiple pathways dominate at different atmospheric heights. Nevertheless, the above
reaction pathway gives some insight into the manner in which ion-neutral chemistry can enhance complex hydrocarbons
at the cost of other carbon-bearing species, such as methane. Acetylene and ethylene may have an interesting connection
to hazes observed in some exoplanets; they are specifically mentioned by \citet{Zahnle2009b} as possibly contributing to 
the hazes observed in HD 189733b and other exoplanets \citep{Pont2008,Bean2010,Demory2011,Sing2011}.}

\sed{Although our model is not directly applicable to irradiated exoplanets, we can speculate what would happen on such planets,
by examining the results of comprehensive photochemical models. \citet{Moses2011}, for example, indicate
that UV photochemistry is efficient at destroying the C$_2$H$_2$ in HD189733b when $p_{\rm gas} \lesssim 10$ $\mu$bar 
(see their Fig. 3). They do find far higher abundances for C$_2$H$_2$ overall, such that its volume fraction at $10^{-8}$
with photodissociation and mixing but without cosmic rays may be $\sim 10^{-20}$. If so, the cosmic-ray enhanced volume
fraction may approach $10^{-11}$ at $p_{\rm gas} = 10^{-8}$ bar. Confirming these speculative results will require a
comprehensive photochemistry self-consistently combined with the cosmic ray chemistry.}

Long polyacetylene radicals are believed to contribute to PAH and other soot
formation both in the interstellar medium \citep{Frenklach1989} and in atmospheres of
e.g. Titan \citep{Wilson2003}. In an atmospheric environment, polyacetylene radicals
tend to react with polyacetylenes to produce longer chains \citep[][their R1, R2]{Wilson2003}.
Further chains can build upon radical sites on these polyacetylenes and may lead to cyclization
of the structure, possibly initiating soot nucleation \citep{Krestinin2000}. Although our network
does not incorporate near this level of complexity, our model does include the constituent parts of 
this process: C$_2$H$_2$ and C$_n$H. These species currently seem to be the most likely candidate 
constituents for PAH growth.

\subsubsection{\sed{C$_{10}$H}}

According to our calculations, one of the most common ions to experience dissociative recombination
in the upper atmosphere is C$_{10}$H$_2^+$. The favored dissociative recombination of this large 
carbon-chain cation in our network is:
\begin{equation}
 {\rm C_{10}H_2^+} + e^- \rightarrow {\rm C_{10}H + H}.
 \label{eqn:carbon-chain}
\end{equation}
The cosmic ray ionization results in an exceptionally high steady-state fractional abundance of 
C$_{10}$H in the very upper atmosphere of exoplanet ($p_{\rm gas} \sim 10^{-6}$), corresponding to a 
density of $n({\rm C_{10}H}) \approx 1000$ cm$^{-3}$, or a fractional abundance of 
$n({\rm C_{10}H})/n_{\rm gas} \sim 10^{-10}$. \sed{It is unlikely that a long polyacetylene radical would be 
so abundant in exoplanetary atmospheres because of the numerous destruction pathways that should exist
in a high-temperature high-density environment for a species so far from thermochemical equilibrium.
Planetary atmospheres are expected to favor aromatic over aliphatic species, and partly for this reason, 
the high abundance predicted by our model nevertheless suggests possibly enhanced abundances of larger 
hydrocarbons built up from reactions between and C$_{10}$H and C$_2$H$_2$.}

\sed{Our model also includes the simple mono-cyclic aromatic hydrocarbon Benzene. We find that cosmic rays enhance
the abundance of Benzene in the upper atmosphere, when $p_{\rm gas} \lesssim 10^{-6}$ bar, by several orders of magnitude.
Since Benzene normally has a very small volume fraction ($< 10^{-50}$) at these low pressures, we assume a quenching
scale height of $p_{\rm gas} = 10^{-3}$ bar, and find that the volume fraction of Benzene still remains well below 
$10^{-20}$ in the absence of cosmic rays, but does achieve an volume fraction of $\sim 10^{-16}$ when cosmic rays are
present. Although this is a significant enhancement, Benzene would be very difficult to observe at this volume fraction.}

\subsubsection{\sed{Chemi-ionization}}

The OSU chemical network used in the \textsc{Nahoon} code also provides us with an opportunity to 
explore purely chemical avenues for enhancing ionization. The single reaction involving two neutral 
species that impacts the ionization is the chemi-ionization reaction:
\begin{equation}
 {\rm O + CH} \rightarrow {\rm HCO^+} + e^{-}.
 \label{eqn:neutral-neutral}
\end{equation}
This reaction has a rate coefficient of  $k = 2 \times 10^{-11}(T/300 \, {\rm K})^{0.44}$ cm$^{3}$
s$^{-1}$,  accurate to within $50\%$ at $T < 1750$ K \citep{MacGregor1973}. \sed{We can use the energetics from \citet{MacGregor1973}
 to construct the reverse reaction, following the method of \citet{Visscher2011}. The reverse reaction is slightly endothermic, with an
activation energy of $\sim 0.4$ eV, or $\Delta E \approx 4600$ K. These reaction taken together will consistently produce a
steady-state degree of ionization, $f_e = n(e^-)/n_{\rm gas} \sim 10^{-11}$ in our model. This degree of ionization is a non-equilibrium steady-state
value, resulting from the enhanced abundance of CH and O predicted by our kinetics model. The impact of this
reaction has not been fully explored and is outside the scope of this paper. In the absence of any significant ionizing 
source, if either chemical quenching or photodissociation enhance the abundances of O and CH significantly, the degree
of ionization would then be enhanced, and ion-neutral chemistry may become important even in these regions.}

\subsubsection{\sed{Ammonia}}
\label{sec:ammonia}

\sed{
In the case of ammonia, we do not invoke quenching at all. The species NH has a reasonable thermochemical volume fraction
in the upper atmosphere \citep{Bilger2013}. Ammonia has a straight-forward connection to the cosmic ray ionization rate via 
the reactions:
\begin{align}
  {\rm H_3^+ + NH} &\rightarrow {\rm NH_2^+ + H_2},\\
 {\rm NH_2^+ + H_2} &\rightarrow {\rm NH_3^+ + H},\\
 {\rm NH_3^+ + H_2} &\rightarrow {\rm NH_4^+ + H},\\
 {\rm NH_4^+} + e^- &\rightarrow {\rm NH_3 + H}.
\end{align}
As such, its volume fraction follows the cosmic ray flux somewhat closely, at least in the upper atmosphere; for higher pressures
nitrogen becomes locked into N$_2$. Our model predicts an abundance of NH$_3$ in the upper atmospheres almost five orders
of magnitude enhanced, bringing the volume fraction from $\sim 10^{-12}$ to $\sim 10^{-7}$. Our model therefore predicts
observable quantities of ammonia in the upper atmospheres of free-floating giant gas planets, in regions where the cosmic rays
are not effectively shielded by the magnetic field. As such, for low-mass substellar objects, this may conceivably be a 
source of variability.}

\section{Conclusions}
\label{sec:conclusion}

The application of \citet{Rimmer2013} to a model atmosphere calculated using \textsc{Drift-Phoenix}
with $\log g = 3$ and $T_{\rm eff} = 1000$ K provides us with an ionization rate. We include the 
ionization rate, as well as the temperature profile, atmospheric number density and elemental abundances 
as input parameters to a time-dependent chemical network. This allows us to calculate the number density 
of ions as well as the abundances of a variety of atomic and molecular species.

The ion-neutral chemistry is responsible for much of the prebiotic chemistry. This preliminary
approach to exoplanet ion-neutral chemistry suggests that it is significant for producing
hydrocarbon chains, and may help to drive PAH production in oxygen-rich atmospheres, possibly giving 
the upper atmospheres of these planets \sed{a chemistry similar to that expected from those objects with an} 
enhanced C/O ratio.  Finally, chemi-ionization processes may also significantly enhance the electron 
fraction in the cloud layer, \sed{if the abundances of atomic oxygen and CH are both above their equilibrium values}. 
The formation of saturated carbon chains is closely connected to the formation of amino acids and other 
important pre-biotic species. \sed{Our results suggest that ion-neutral chemistry has a role to play in
hydrogen-dominated environments at altitudes of $\sim 1$ $\mu$bar, and seems, at least without fast mixing
and UV photochemistry present in atmospheres of irradiated planets, to generally enhance the abundances
of some complex hydrocarbons, some of which may be relevant to prebiotic chemistry.}

This work on a free-floating exoplanet may also be applicable to the directly imaged exoplanets 
orbiting HR 8799, but in order \sed{to explore this mechanism for the haze} in the upper atmosphere of the exoplanet 
HD 189733 b we need to model the atmospheric chemistry for an irradiated exoplanet. It will then be 
important to account for the effect of the magnetic field of the host star as well as the planet
upon cosmic ray transport. \sed{The effect of stellar winds and their interaction with the magnetic field of the giant
gas planet may also become important \citep[see, e.g.][]{Vidotto2012}.} It will also be necessary for ion-neutral chemical 
models to provide absolute abundances instead of ratios. Although the result of this work has \sed{indicated}
significant trends, and \sed{suggests} that ion-neutral chemistry \sed{may be} important part of the atmospheric 
chemistry of free-floating giant gas planets, much work still needs to be done to develop a useful 
ion-neutral gas-phase network appropriate for the atmospheres of giant gas planets, both free-floating 
and the hot Jupiters.

\renewcommand{\abstractname}{Acknowledgements}
\begin{abstract}
We highlight financial support of the European Community under the FP7 by an ERC starting grant. 
We thank Peter Woitke for his help with determining time-scales to chemical equilibrium. We also thank the
anonymous referees for comments that have significantly improved the quality of this paper. 
This research has made use of NASA's Astrophysics Data System. We thank Ian Taylor for his technical support.
\end{abstract}

\bibliographystyle{apalike}
\bibliography{master,master-2} 

\end{document}